**Resistive switching and long-range filaments in metal/DMSO liquid systems for three-dimensional, multi-terminal connection schemes with on demand dynamic reconfigurability.**

*Roshani Madurawala, Kerstin Meurisch, Louis Joswig, Anna Lina Wyschkon, Maik-Ivo Terasa, Sören Kaps\*, Alexander Vahl and Rainer Adelung\**

R. Madurawala, K. Meurisch, L. Joswig, A. L. Wyschkon, M-I. Terasa, S. Kaps, R. Adelung
Department of Materials Science – Chair for Functional Nanomaterials, Faculty of Engineering, Kiel University, Kaiserstraße 2, D-24143 Kiel, Germany
E-mail: ska@tf.uni-kiel.de

A. Vahl
Department of Materials Science – Chair for Multicomponent Materials, Faculty of Engineering, Kiel University, Kaiserstraße 2, D-24143 Kiel, Germany
Kiel Nano Surface and Interface Science KiNSIS, Kiel University, Christian-Albrechts-Platz 4, D-24118 Kiel, Germany
Leibniz Institute for Plasma Science and Technology, Felix-Hausdorff-Straße 2, 17489 Greifswald, Germany

Funding: German Research Foundation (DFG) in the Collaborative Research Center SFB 1461 (Project-ID 434434223).

The human brain, with its energy-efficient and massively parallel architecture seamlessly integrates memory and computation. Its topology and functionality serve as the inspiration for the field of neuromorphic computing. Realizing brain-like hardware requires the integration of fundamental properties such as synaptic plasticity, self-organization, hierarchical and modular structures, as well as three-dimensional connectivity. Current challenges lie in developing liquid based neuromorphic material systems with facile fabrication, three-dimensional processing, and brain-like conductivity. This work presents ionotronic systems – i.e., systems that incorporate the movement of both electrons and ions - to obtain dynamically reconfigurable conductive filaments. Our method employs an electrolyte where an anode reservoir produces ions in-situ, enabling electrode-dependent tunability and sustained operation without ion depletion. This manuscript presents four ionotronic systems. Each system grows brain inspired three-dimensional wires contacting two or more electrodes exhibiting resistive switching at connection on a micrometer scale as well as a nanometer scale, demonstrating hierarchical organization and functionality. Furthermore, these conducting filaments are capable of being disrupted by an external electric field or dissolved over time in the ionotronic system, emulating blooming and pruning aspects of plasticity.

**1. Introduction**

The human brain is a massively parallel and energy-efficient system where memory and computation are tightly integrated across networks of neurons and synapses; the structural and functional map of those connections is often referred to as the connectome.[1–3] The morphology and the function of this network provide the foundation for *neuromorphic computing*. Neuromorphic computing seeks to emulate brain-inspired computational principles in software and hardware, e.g., by using spiking networks, local plasticity rules, and architectures that decentralize memory and processing.[4–7] By mimicking neuronal and synaptic behavior, neuromorphic approaches enable aspects of in-memory computing and event-driven processing that can substantially reduce energy consumption of computation offering a promising path to overcome the limitations of conventional computer architectures.[8–11]

Developing the next generation of brain-inspired computing systems therefore motivates efforts to build brain-inspired hardware concepts. This endeavor requires simultaneous consideration of several key principles, including distributed and local plasticity (dynamic reconfigurability), operation near criticality, self-organization, three-dimensional wiring and processing, robustness, hierarchy and modularity, massive parallelism, nonlinearity, scalability, and sparsity. No single hardware platform today fully realizes all of these features at biological scale; instead, current research explores tradeoffs and seeks to combine multiple mechanisms.[7,12]

A variety of approaches investigate and optimize individual key features. One example is dynamic connectivity, which is a prerequisite for distributed plasticity and enables learning and adaptation.[13–15] Dynamic connectivity is the ability of a system to adaptively modulate the presence of a connection as well as the strength of a connection in response to stimuli. Such reconfigurability can be implemented by resistive switching, either gradual modulation of conductance (analog switching) or through formation and rupture of conductive filaments as observed in many memristive devices.[16–19] Ion-mediated and mixed ionic-electronic devices, i.e ionotronic devices, are also promising because ionic motion adds rich temporal and chemical dynamics that can emulate biological synaptic behavior. These material systems offer distinct trade-offs in terms of analog resolution, retention, endurance, and fabrication scalability.[14,20,21]

An example for utilizing mixed ionic electronic conductivity is brain-inspired organic electronics where conducting polymers are utilized to fabricate ionotronic devices.[22] Recent developments in liquid-based neuromorphic device concepts closely facilitate neuro-emulating hardware. Examples of such concepts include liquid-based transistors and memristors, most of which are confined to two-dimensional architectures and rely on clean-room fabrication techniques that pose scalability challenges, while constrained by limited ionic species in the electrolyte.[23–25]

Beyond dynamic connectivity, adopting three-dimensional network architectures is an essential key feature, as it allows the formation of complex networks with a greater number of nearest neighbors, enhancing information processing capabilities beyond what is achievable in purely two-dimensional designs.[26,27] For example, three-dimensional topologies focus on increasing the rate of information processing, rather than increasing the density of transistors on a chip.[9] By employing a three-dimensional spatial embedding, networks achieve both efficiency through sparsity, wiring-cost minimization, and global integration—enhancing computational capacity beyond what flat architectures can offer. Another key feature is reconfigurable connectivity, characterized by electrical connections whose formation, dissolution, or temporary inactivity are stimulus-dependent. This enables energy-efficient operation by preventing redundant pathways.[28–31]

In this work, we present a concept in which ionic dynamics and redox reactions facilitate the self-organized formation and dissolution of conductive filaments without the need for cleanroom fabrication. By employing in-situ ion generation, the system overcomes the inherent limitations of finite ionic reservoirs in liquid environments, enabling more robust and controllable filament behavior. While conventional memristive and filamentary devices demonstrate resistive switching and synaptic plasticity, they are often limited by two-dimensional architectures or by constraints on ionic reservoirs.[31] Here, we propose an alternative approach that leverages in-situ ion generation and three-dimensional connectivity to realize dynamically reconfigurable filaments. Furthermore, these filaments exhibit resistive switching on two hierarchical scales: (1) on the micrometer scale, where a rapid increase in current results from the formation of a conductive filament emulating axonal blooming; and (2) on a nanometer scale, where the formed conductive filament has the potential to exhibit resistive switching cycles emulating synaptic plasticity. The dynamic network exhibits self-organization due to the diffusion limited aggregation process taking place in the initial stages of filament growth. This manuscript therefore elucidates the formation of reconfigurable, three-dimensional, controlled-self-organized long-range connections formed via electrical stimulation of multi-terminal active electrode/DMSO/active electrode, active electrode/DMSO/inert electrode, active electrode/1 M HCl/active electrode, and active electrode/1M HCl/inert electrode arrangements.

## 2. Results and Discussion

This section is categorized into four sub-sections to navigate the topology and properties of the reconfigurable filaments. The goal of this manuscript is to present the topology and functionality of the filament in a micrometer scale for the broader perspective of emulating selected features of synaptic plasticity such as growing and regressing of a conducting filament. First, the growth of a single filament between two contacts in the metal/DMSO system is demonstrated. Then the correlation between the factors effecting the filament growth and the properties of the resultant filament is discussed. This consideration is significant when aiming to tune the filament or to correlate the electrical stimuli to the topology of the filament. Secondly, the electrical properties of the grown filament are analyzed by applying a series of cyclic voltage sweeps between -4 V to 4 V to obtain a resistive switching phenomenon without contributing to the growth of the filament. Thirdly, multi terminal arrangements are studied, with a particular interest in exploring to which extent sequential growth of connections can be achieved between multiple electrodes in the third dimension. Fourthly, the growth and dissolution of the filament are investigated at the example of a metal/1 M HCl system to controllably manipulate the dynamic equilibrium between formation and dissolution of long-range connections.

## 2.1. Growth of a single reconfigurable filament in DMSO

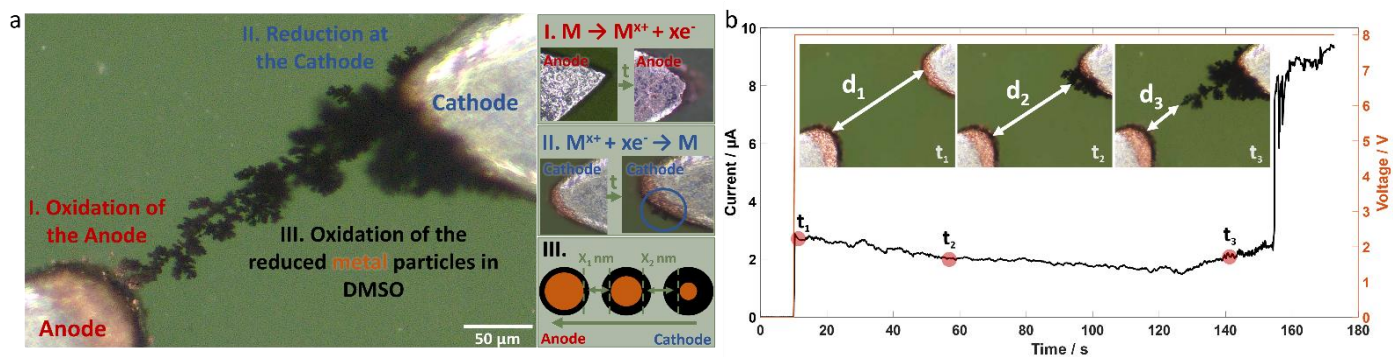

**Figure 1** Filament growth process. (a) Light microscopic image of the electrodes on the green colored PCB, labeled anode (red) and cathode (blue), and the grown filament (black). The schematics on the right a I. and a II. represent the anodic and cathodic reactions responsible for the filament growth and a III. shows the simultaneously occurring partial oxidation of the filament. Since the filament growth starts at the cathode, the oxide shell is the thickest nearer to the cathode (schematic a III). (b) Current profile measured over time during filament growth corresponding to the light microscope image shown in **Figure 1a**. An abrupt increase in current is visible close to 160 seconds, which corresponds to a connection being made between the two electrodes. $t_1$, $t_2$, and $t_3$ are time points visualizing the growth stages during the filament growth process.

In **Figure 1a**, an optical microscopy image of a PCB with a 5 µl DMSO droplet is shown. After a voltage of 8 V has been applied between the anode (**Figure 1a** bottom left labeled in red) and the cathode (**Figure 1a** top right labeled in blue), the contrast in the area in between the electrodes has changed and dark areas have formed (**Figure 1a** black). The formation sequence of these dark areas can be seen in the insets of **Figure 1b** from $t_1$ to $t_3$. These areas correspond to the filament discussed in this manuscript. The mechanism for the formation of this filament is proposed below.

DMSO is a polar aprotic solvent. When a droplet of DMSO is transferred onto the electrodes, DMSO molecules are physiosorbed onto the surface forming a protective layer.[32] However, under the electric field the anodic material, M, undergoes oxidation releasing metal ions into the electrolyte which results in an ionotronic system. This is evident by the dissolution of the anode shown in **Figure 1a** (schematic I.) where the respective reaction is given in **Equation 1**.

$$nM \rightleftharpoons M^{n+} + ne^- \qquad\qquad 1$$

The release of the metal ions at the anode facilitates a concentration gradient between the electrodes in addition to the electric field, therefore, ion diffusion as well as ion drift occur in the liquid droplet. At the cathode the metal ions undergo reduction and deposits on the cathode. It can be observed that the filament grows from the cathode to the anode corresponding to consecutive metal reduction at the shortest distance from the anode ($d_1$, $d_2$ and $d_3$ in **Figure 1b** inset). Closer to the cathode, the diffusion limited aggregation is quite prominent and the filament shows fractal growth away from the cathode. Fractal growth is a repetition of a growth pattern which demonstrates hierarchy in a shorter and longer length scale. There is a high projected surface coverage of the filament at the cathode compared to the anode, which is a result of diffusion limited aggregation. As the filament approaches the anode the distance between the region of oxidation and region of reduction reduces (compare $d_1$ to $d_3$ in **Figure 1b**), resulting in an increase of dominance in electric field driven ion migration, hence making the growth process faster and more directional towards the tip of the anode.

The interplay between diffusion and electric field driven ion migration determines the self-organization of the filament. The degree of self-organization and the directed growth can be controlled by altering the applied voltage for a given electrode distance. When voltages ranging from 7 V to 13 V were applied between two electrodes separated with a distance of 260 µm, it was observed that the time taken to connect the two electrodes with the filament, decreases with increasing applied voltage. Additionally, the $I_{on}/I_{off}$ ratio increases with increasing voltage. The $I_{on}/I_{off}$ ratio is calculated with the current output before and after the resistive switching phenomena occurring when the filament connects the two electrodes. The resistive switching here refers to the rapid current increase shown in **Figure 1b** which occurs when the filament connects the anode and the cathode in the micrometer scale. This is visualized in **Figure S1** in the supporting information. The distance between the electrodes is kept constant. With increasing voltage, the filament gets comparatively thinner and more directional, which is evident by the light microscopic images in **Figure S1**.

From the top planar optical view of the filament, an approximation for the fractal dimension, a space filling factor, can be calculated. In theory, the fractal dimension value of 1 corresponds to a smooth line.[33] It was observed that for voltages from 7 V to 13 V the fractal dimension is inversely related to the voltage. Which signifies that, higher the voltage, the more directional the filament grows. This relationship is visualized in supporting **Figure S2**. Furthermore, the $I_{on}/I_{off}$ ratios and the fractal dimension has an inverse relationship, indicating that the switching of the resistance is more prominent when the filament grows more directional.

Analogous to the key dimensions for brain-inspired hardware development mentioned above, this single filament growth demonstrates self-organization near criticality as a balance between disorder and order in terms of diffusion limitation and electric field driven growth. In other words: The filament grows without a strict blueprint, however, shows directionality due to the concentration gradient, electric field and the positioning of the electrodes.

When the filament connects the two electrodes there is an abrupt increase in the output current, which corresponds to the switching of the resistance in the micrometer scale. However, the output current of the connected filament is in the µA range. The **Table 1** in the supporting information elaborates the current at connection values from 7 V to 13 V.

In the current versus time plot depicted in **Figure 1b**, when the connection was formed the current reached 9 µA. This signifies that the filament is not completely metallic. The proposed hypothesis is that metal oxides are present in the filament. Optical observations suggest that the filament is black in color. Considering the anode material from **Figure S3**, the filament is proposed to be copper oxide as well as tin oxide. For the scope of this manuscript metal oxides will be collectively referred to as $M_yO_x$. During the growth of the filament, as the metal ions deposit at the cathode, the particles or agglomerates react with the dissolved water in DMSO and creates $M/M_yO_x$ core-shell particles (see **Figure 2b**) following the **Equation 2**.[34]

$$yM + xH_2O \rightleftharpoons M_yO_x + 2xH^+ + 2xe^- \qquad 2$$

For a fixed electrode distance and applied voltage, a faster filament growth leads to a more directed connection and results in a higher current once the connection is formed. This can be seen from the values in **Table T1** of the supporting information at higher voltages. For example, consider two experiments conducted at 11 V with identical parameters. The filament that formed a connection after 19 s conducted 17.66 µA, whereas the one that required 40 s to connect conducted only 6.76 µA. This is due to the formation of the oxide shell over time, reducing the conductivity of the filament (Supporting information **Table T2**). With time, the oxide shell gets thicker increasing the distance between the metal cores (compare $X_1$ and $X_2$ in **Figure 2b**)

In summary, these results indicate that a conducting filament can be formed, or a connection can be established between two electrodes in a DMSO medium by a redox reaction. The growth of this connection is reconfigurable so that by altering the applied electrical field the topology as well as the resistance of the filament could be altered. Reconfigurable network formation is a key factor when emulating the plasticity of the brain. Synaptic plasticity in the brain is exhibited structurally as well as functionally either by reconfigurations in the physical structure of the synaptic connections or by changing their synaptic weight or resistance.[15] These changes in the resistance can be utilized to store information. An example for one such utilization is the phenomenon known as resistive switching in memritive devices.[35–37]

## 2.2. Resistive switching of the filament formed in DMSO

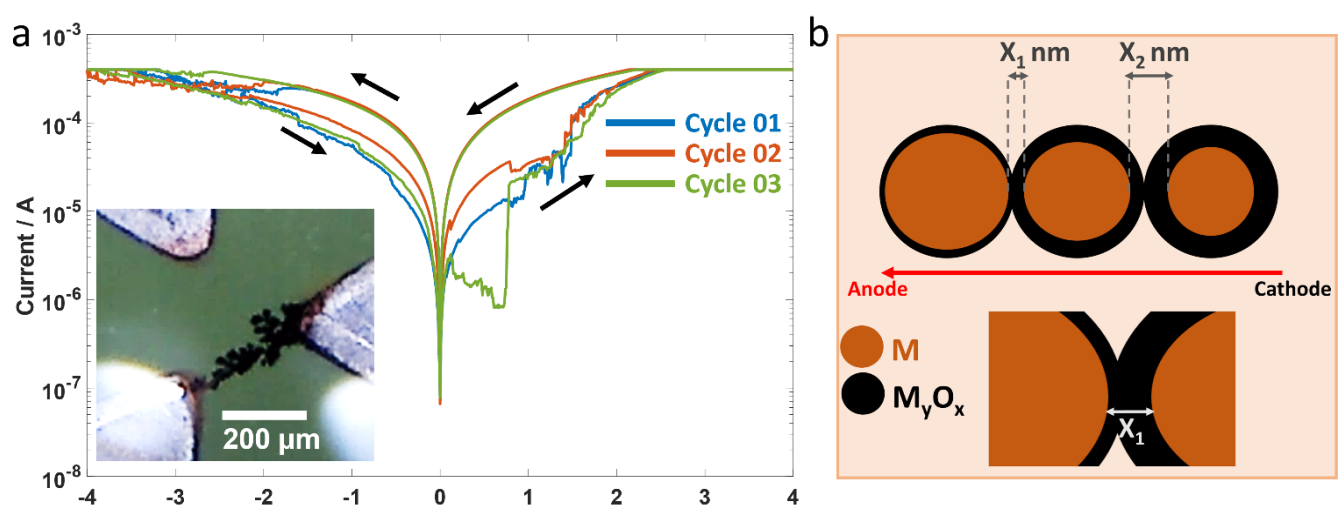

**Figure 2** Resistive switching behavior of the grown filament. (a) I-V plot (logarithmic) of the connected filament when a voltage sweep from -4 V to +4 V was applied between the two connected electrodes. The inset to the left shows the light microscopic image of the corresponding filament. This filament was grown in DMSO at 5 V. (b) Schematic (not to scale) of how a metal-insulator-metal system could form inside the filament.

It was discussed in the previous section that the grown filament exhibits varying conductivity due to the formation of an oxide shell around the metal particles. With time, the oxide shell gets thicker increasing the distance between the metallic cores (compare $X_1$ and $X_2$ in **Figure 2b**). When a voltage sweep was carried out from 0 V to +4 V to -4 V to 0 V, hysteresis loops as shown in **Figure 2a** was observed. The microscopic image in **Figure 3a** shows the corresponding filament in black for the obtained hysteresis behavior.

Metal oxides have been considered as an insulator material in metal-insulator-metal resistive switching (RS) devices.[19,38] Oftentimes there is an active metal and an inert metal on either side of the insulator material. When sufficient electric field is applied, the active metal oxidizes to create cations which migrate towards the inert cathode to undergo reduction, eventually creating a conducting pathway due to electrochemical metallization (ECM). However, oxides, specifically transition metal oxides such as CuO contain oxygen defects/vacancies. In the case of CuO, similar to ZnO and $TiO_2$ RS devices, there is an electrically driven migration of oxygen vacancies.[39–41] These positively charged oxygen vacancies are created due to the negative charge of oxygen ions. In an electric field, these oxygen vacancies can move to the cathode while generating a negatively charged agglomeration of oxygen ions close to the anode. As a result, positively charged oxygen vacancies move to the cathode eventually creating a conducting pathway. There can also be local heating phenomena on a nm scale such as Joule heating which contributes to the reduction of the resistance by providing thermal energy to facilitate the migration of ions.[35] When the polarity of the electric field is reversed the conducting particles diffuse in the insulator material breaking the conducting pathway due to repulsion of the charged species which then leads to the high resistive state. In this nanoscopic metal-insulator-metal system, when a voltage is applied it can be assumed that the positively charged particles migrate towards the respective cathode. These charged particles would thus lower the resistance between the anode and the cathode collectively resulting in the low resistive state and vice versa.

Both processes, the growth of the microscopic filament between the two triangular metal electrodes as well as the nanoscopic conducting pathway between the two metal cores, show topological hierarchy in the system in terms of controlled self-organization. This switching behavior occurs due to the migration of charged particles under an electric field in a nanometer scale inside the micrometer scale filament. This RS phenomena contributes to the multiple scales of functional hierarchy in the system. In addition to the emulation of axonal growth on a µm scale, these reconfigurable filaments hold the potential of exhibiting varying resistive states. The research into the electrical properties of these filaments is at its initial stages. However, these initial results show promising potential in applications of neuromorphic hardware. The next step of such reconfigurable individual filaments is the exploration of their ability to form a three-dimensional network.

## 2.3. Filament growth in three-dimensional multi-terminal arrangement in DMSO

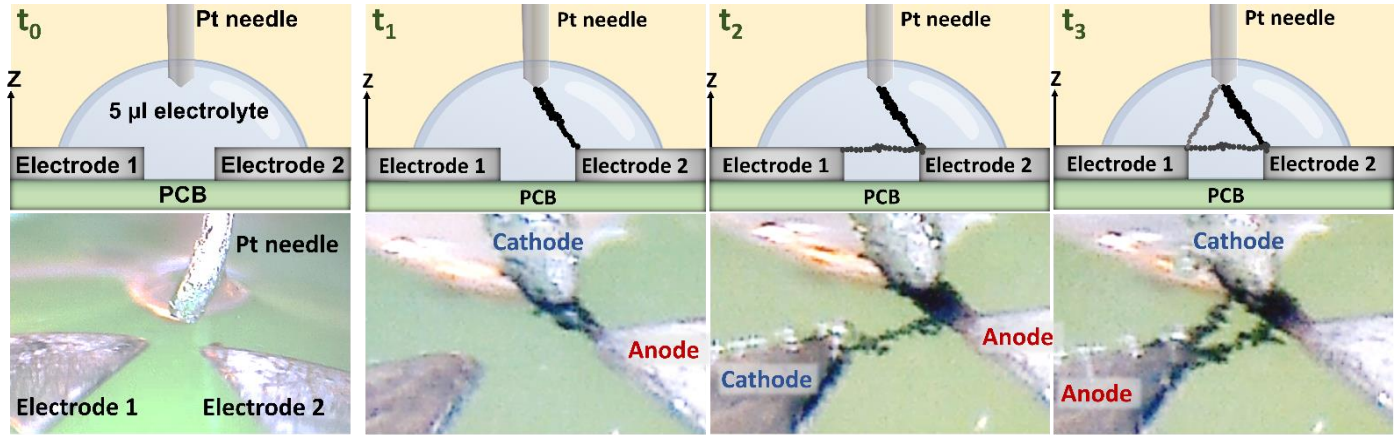

**Figure 3** Filament growth in the z axis, against gravity. The platinum (Pt) needle was introduced from the top, perpendicular to the green PCB with the three triangular electrodes. At $t_0$, no voltage was applied. The schematic of the setup is illustrated above the respective light microscopic image. At $t_1$, 14 V was applied between the Pt needle and electrode 2 which is the anode, until a filament was formed from the cathode to the anode (electrode 2 in $t_1$). At $t_2$, 8 V was applied between electrode 1 and 2 to grow a second filament without disrupting the existing first filament. Finally, a third filament was grown by applying 14 V between the Pt needle and the anode (electrode 1 at $t_3$).

The structure-function relationship of the human brain works across different scales which relates to its hierarchy and modularity. The structural aspect however, composes of the complex anatomical network of neurons which is in a three-dimensional space.[27,42] When emulating the properties of the brain it is important that these reconfigurable filaments could form networks as well as connect individual electrodes without disruption. **Figure 3** shows a series of schematics and corresponding microscopic images of growing a simple network arrangement in the z-axis with three electrodes and three connections. For these experiments, an external Pt needle was employed from the top (Z axis) as shown in **Figure 3** $t_0$. When forming the first filament shown in the second schematic ($t_1$) of **Figure 3**, the anode undergoes oxidation and the ions move against gravity towards the Pt needle where the ions are reduced. The second filament is formed between the two triangular electrodes without disrupting the first connection ($t_2$). Due to the oxidation of the filament over time, short circuiting the Pt needle and electrode 1 is possible even though they are connected via two filaments, which gives rise to a conducting third filament. This exhibits that these reconfigurable filaments have the potential of forming three-dimensional networks. Furthermore, these filaments could be formed individually between multiple electrodes, depending on the number of electrodes in the system. Furthermore, the filament formation is not limited by the particle concentration in the liquid. This is the advantage of employing an anodic dissolution process to introduce the particles that form the filament.

The key finding in this section lays the foundation to fully dynamic, three-dimensional, robust connections and architectures. When there is no voltage applied to the filament, it undergoes oxidation depending on the thickness of the filament, becoming electrically insignificant to further filament formation. Filament thickness increases with growth time, enhancing conductivity in a manner analogous to synaptic weight modulation. For example, a filament grown at 10 V until reaching 1 mA exhibited a stable resistance of 1.6 k$\Omega$ over 20 voltage sweeps within 1 hour. 50 resistance values measured during cyclic voltage sweeps over 3 hours after the growth of the filament are plotted and elaborated in **Figure S4**.

When it comes to plasticity, reorganization of the network blooming and pruning are very important factors for network efficiency and sparsity. In these dynamic networks, pruning or the aspect of forgetting in the brain can be achieved either by the help of an electric field or by chemical dissolution. The advantage of in-situ ion generation is highlighted with the six-electrode system presented in **Figure S5**. Here the sequential formation of individual filaments without refreshing the electrolyte is shown and when there is no potential applied the connections undergo rupture. **Figure S6** depicts the formation of connections between three electrodes in two different aspects. First aspect is the divergent contacting of the electrodes simultaneously. Second aspect is the three electrodes contacting divergently in consequent steps. **Figure S7** shows the filament formation, breakage and reformation under the influence of an electric field. Here four electrodes are employed in the system where the filament is formed between two electrodes and the remaining two electrodes are used to manipulate the filament accordingly. In summary, the supporting information provide evidence for filament regression by breakage due to an application of an electric field, spontaneous rupture at inactivity, and gradual inactivity due to filament oxidation. Furthermore, the dissolution of the filament was further investigated in DMSO by adding HCl to chemically etch away the metal oxide. The light microscopic images before and after the chemical dissolution is elaborated in **Figure S8**.

## 2.4. Formation and dissolution of the network connections in 1 M HCl

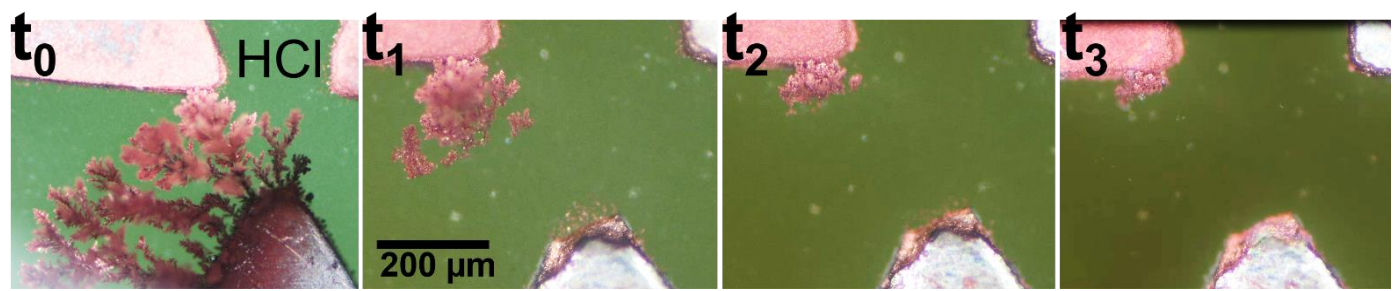

**Figure 4** Filament growth and dissolution in HCl. Microscopic images from $t_0$ to $t_3$ show the filament growth and dissolution in HCl. Image $t_0$ show the grown filament in HCl at room temperature when 0.6 V was applied. The current versus time profile of this filament growth can be found in supporting information **Figure S9**. The filament dissolution time taken from $t_1$ to $t_3$ is 140 seconds when no voltage was applied in aqueous HCl.

As discussed in the previous sections the reconfigurable networks grown in DMSO contain metal oxides. The metal oxide dissolution of the filament the in HCl can be expressed according to the **Equation 3**.[43] Respective microscopic images and a current-time-plot can be found in supporting information **Figure S9**.

$$MO + 2xHCl \rightarrow MCl_x + xH_2O \qquad 3$$

In addition to DMSO electrolyte, the growth of a conducting filament was obtained in 1 M aqueous HCl solution. This growth in HCl could also be observed in the vertical plane as shown in the supporting information **Figure S10**. The **Figure 4** $t_0$ shows two electrodes with multiple branched filaments in between the electrodes. Optically, the filament contains black areas close to the cathode and copper-colored branches, dominantly close to the anode. The growth time of this connection was 5.4 minutes at 0.6 V and the final readout current was 830 µA (current and voltage versus time plot in supporting information **Figure S9**). **Figure 4** therefore, represent two distinct phenomena: First phenomenon is the initial redox reaction that occurs at 0.6 V which is the electro-dissolution of the anode in 1 M Cl- media and consequent formation of the filament seen in $t_0$. The second phenomenon occurs at 0 V, in **Figure 4** from $t_1$ to $t_3$. Here, the previously formed filament undergoes spontaneous-dissolution. However, the dissolution of the filament in HCl can be observed at low voltages such as 50 mV as depicted in **Figure S10**. This indicates the capability to utilize small readout voltages without impeding the dissolution mechanism.

Considering the first phenomenon, initially, the system is at a high $Cl^-$ concentration. At higher chloride concentrations, anodic reactions analogous to that in **Equation 4** may occur, with the increased availability of chloride ions further promoting M–Cl compound formation.

$$M + Cl^- \rightarrow MCl + e^- \qquad\qquad 4$$

Continuous oxidation of the anode changes the concentrations of the ionic species in the M-$Cl$-$H_2O$ ionotronic system. At the cathode electro-deposition and diffusion limited aggregation of the metal from a chloride solution takes place. In addition to the reduction of the metal at the cathode, precipitations of $M_yCl_x$ can take place at free $Cl^-$ concentrations below 1 M.[44]

The second phenomenon is the dissolution of the formed filament. Some metals can undergo corrosion according to **Equation 4**, and metal oxides dissolve in acids such as HCl according to **Equation 3**. In the M-$Cl$-$H_2O$ ionotronic system these species can form a number of compounds that are soluble in an aqueous medium.[43–45]

The key finding of this section is the dissolution of the filament. This facilitates the regression or the pruning part of plasticity. By changing the electrolyte to aqueous HCl a less directed connection can be formed between two or more electrodes with only 500 mV and at the absence of the electrical stimuli or at a very low stimuli the filament is dissolved in the electrolyte. This demonstrates features of neuroplasticity in the system. The ionic composition and the microscopic and nanoscopic dynamics are complex and opens potential research avenues. Additionally, the system exhibits emergent collective brain-like behavior that can be harnessed for functional applications.

## 3. Conclusion

This work demonstrates reconfigurable filaments in an ionotronic liquid medium that can enable hardware systems with brain-like topology and functionality. By harnessing in-situ ion generation from an electrode reservoir, our method overcomes concentration limitations of conventional electrolytes. Furthermore, this method enables electric field driven, three-dimensional self-organization of conductive pathways. The resulting filaments are not only tunable in their growth kinetics and electrical properties, they can also be deactivated, ruptured, or dissolved in a manner analogous to brain-like plasticity. Importantly, the observation of hierarchical aspect of long-range and short-range resistive switching within these filaments' points toward the potential for memory operations, through further characterization and optimization. Collectively, these filaments suggest a pathway toward scalable, energy-efficient neuromorphic hardware, where network topology and functionality can be adaptively reconfigures in a liquid medium. Looking ahead, investigating stability, switching dynamics, and large-scale integration will be essential to advancing this approach toward practical in-memory computing architectures with the added advantage of having a liquid phase electrolyte.

## 4. Experimental Section/Methods

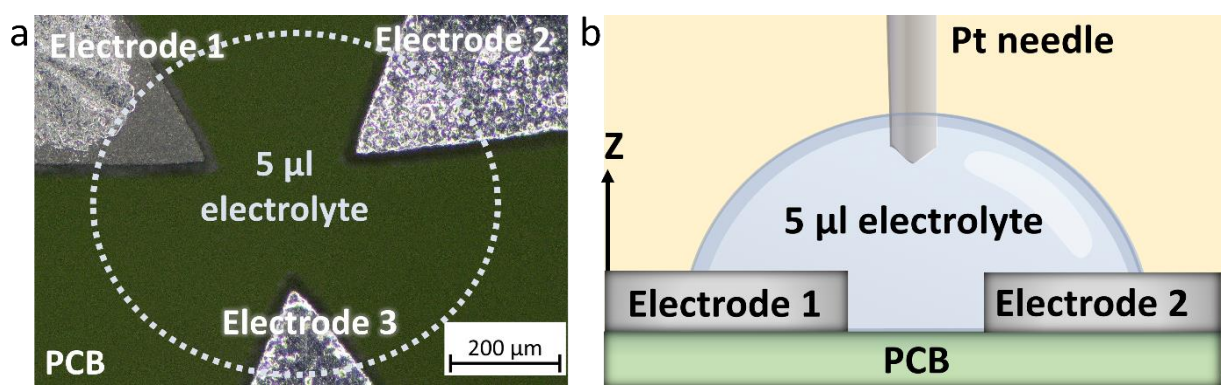

**Figure 5.** Microscopic image and a schematic of the experimental setup. a) The light microscope shows the top view of the experimental setup. The set-up consists of three triangular shaped electrodes on a printed circuit board (PCB) shown in green. The tips of the electrodes have a distance of 260 µm to each other with outlets on each electrode to connect to a source meter. b) The schematic of the side view of the electrode system when an additional Pt (Platinum) needle was manually incorporated from the top for the three-dimensionality measurements as a fourth inert electrode (see **Figure 3**).

### 4.1 Chemicals

The electrolytes used in the experiments were Dimethyl sulfoxide, HPLC grade, 99.9+% (DMSO) provided by Thermo Fisher Scientific and Hydrochloric acid (37 %, CAS: 7647-01-0) from Carl Roth GmbH. In all experiments a 5 µl droplet of electrolyte was placed on the electrodes so that all electrodes were immersed in the electrolyte.

**4.2 Experimental setup**

The experimental setup consisted of a self-designed printed circuit board (PCB) ordered from AISLER Germany GmbH. A single PCB included 16 sets of electrodes. Each set of electrodes contained three electrodes separated with a distance of 260 µm as shown in Figure 01. The electrodes were coated with a protective layer containing Sn as evident from EDX measurements (Supporting information **Figure S3**) and the electrode material was copper. Therefore, for the scope of this work all ionic species involved in the ionotronic system would be named as 'metal' (M). Each single electrode was coupled with a receptacle that could be individually connected to a source meter. The source meters used in all experiments were Keithley 2450 Source Measure Units (SMU). Both current and voltage were monitored during the experiments and were measured with the help of the SMU, controlled using a self-developed custom-made software integrated with a USB microscope and light microscope (Axioscope 5).

For each experiment two of the three electrodes were connected to the SMU. The electrodes were cleaned with deionized water and isopropanol (99.5 % purity, Sigma Aldrich) before each experiment and pat dried so that there was no residue liquid on the electrodes. A constant voltage was applied and the output current was measured. To investigate the filament growth in the third dimension, a Pt needle was incorporated manually from the top as sketched in **Figures 3 and 5**.

**Acknowledgements**

The authors acknowledge funding from the German Research Foundation (DFG) in the Collaborative Research Center SFB 1461 (Project-ID 434434223) and the research unit FOR 2093. The authors also thank Philipp Schadte for the EDX measurements.

# Supporting Information

**Resistive switching and long-range filaments in metal/DMSO liquid systems for three-dimensional, multi-terminal connection schemes with on demand dynamic reconfigurability.**

*Roshani Madurawala, Kerstin Meurisch, Louis Joswig, Anna Lina Wyschkon, Maik-Ivo Terasa, Sören Kaps\*, Alexander Vahl and Rainer Adelung\**

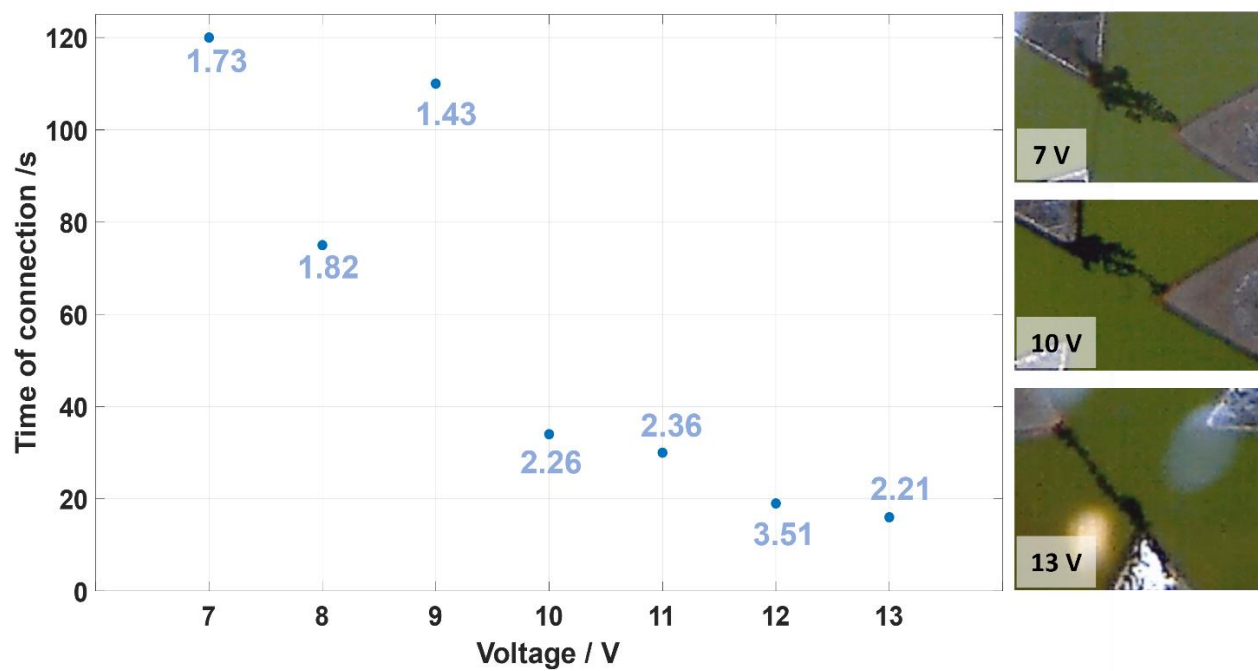

**Figure S1: Time of connection versus Voltage plot with $I_{on}/I_{off}$ ratios at connection**. Each data point is the average of 3 experiments (**Table T1**). The three microscopic images on the right show the connected filament for voltages 7 V, 10 V, and 13 V. Higher voltages result in thinner filaments and a lower time of connection with a higher current at connection. The optical appearance of the filament further suggests that at lower voltages the diffusion limitation dominates for a longer fraction of the filament growth time resulting in more oxidation of the filament.

**Table T1**: **Filament growth t different voltages**. The table shows three experiments conducted for each voltage. Due to the oxidation of the anode, a new set of electrodes were chosen for each voltage. For each experiment a liquid droplet of 5 µl was used. The table shows the time for connection in seconds which is the time from the start of the voltage application until a connection was made. The current at connection was recorded at the 'time for connection'.

| Voltage | Experiment number | Time for connection (seconds) | Current at connection (µA) |
|---|---|---|---|
| 7 | 1 | 121 | 5.12 |
|  | 2 | 128 | 5.17 |
|  | 3 | 109 | 5.16 |
|  | *Average* | *120* | *5.2* |
| 8 | 1 | 94 | 5.19 |
|  | 2 | 67 | 7.75 |
|  | 3 | 64 | 4.78 |
|  | *Average* | *75* | *5.9* |
| 9 | 1 | 91 | 6.71 |
|  | 2 | 129 | 4.81 |
|  | 3 | 110 | 3.42 |
|  | *Average* | *110* | *5.0* |
| 10 | 1 | 32 | 10.47 |
|  | 2 | 28 | 12.17 |
|  | 3 | 41 | 10.62 |
|  | *Average* | *34* | *11.1* |
| 11 | 1 | 19 | 17.66 |
|  | 2 | 32 | 14.62 |
|  | 3 | 40 | 6.76 |
|  | *Average* | *30* | *13.0* |
| 12 | 1 | 15 | 16.04 |
|  | 2 | 16 | 15.54 |
|  | 3 | 25 | 22.97 |
|  | *Average* | *19* | *18.2* |
| 13 | 1 | 15 | 13.30 |
|  | 2 | 16 | 13.79 |
|  | 3 | 15 | 13.89 |
|  | *Average* | *16* | *13.7* |

**Table T2**: **Final filament of three experiments for different voltages**. Three experiments were carried out for each voltage. For each set of voltages, a new electrode set was used and all other parameters were kept constant. For 10 V set time taken for connection for experiment 1,2, and 3 is respectively 32 s, 28 s, and 41 s. For a given voltage and similar conditions, the filaments could exhibit variations. However, in the midst of such variations, certain patterns could be extracted such as: relationship between voltage and current at connection, and fractal dimension and relationship between growth time and current at connection.

|  | 7 V | 10 V | 11 V | 12 V |
|---|---|---|---|---|
| **Experiment 1** | 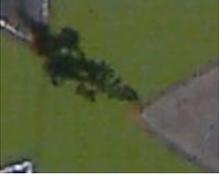 | 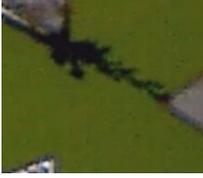 | 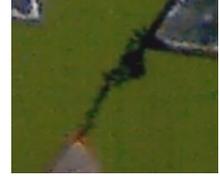 | 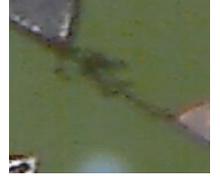 |
| **Experiment 2** | 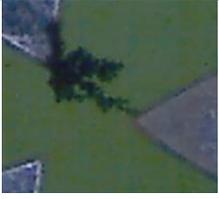 | 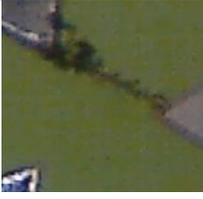 | 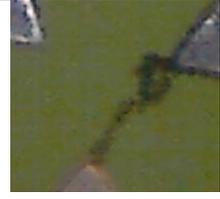 | 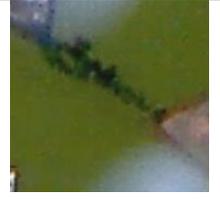 |
| **Experiment 3** | 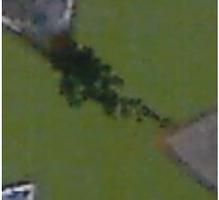 | 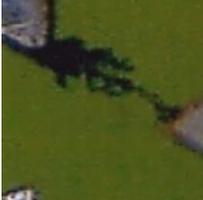 | 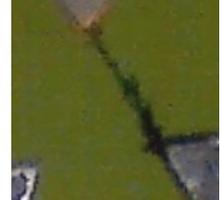 | 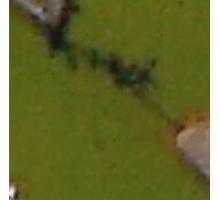 |
| $I_{on}/I_{off}$ | 1.73 | 2.26 | 2.36 | 3.51 |

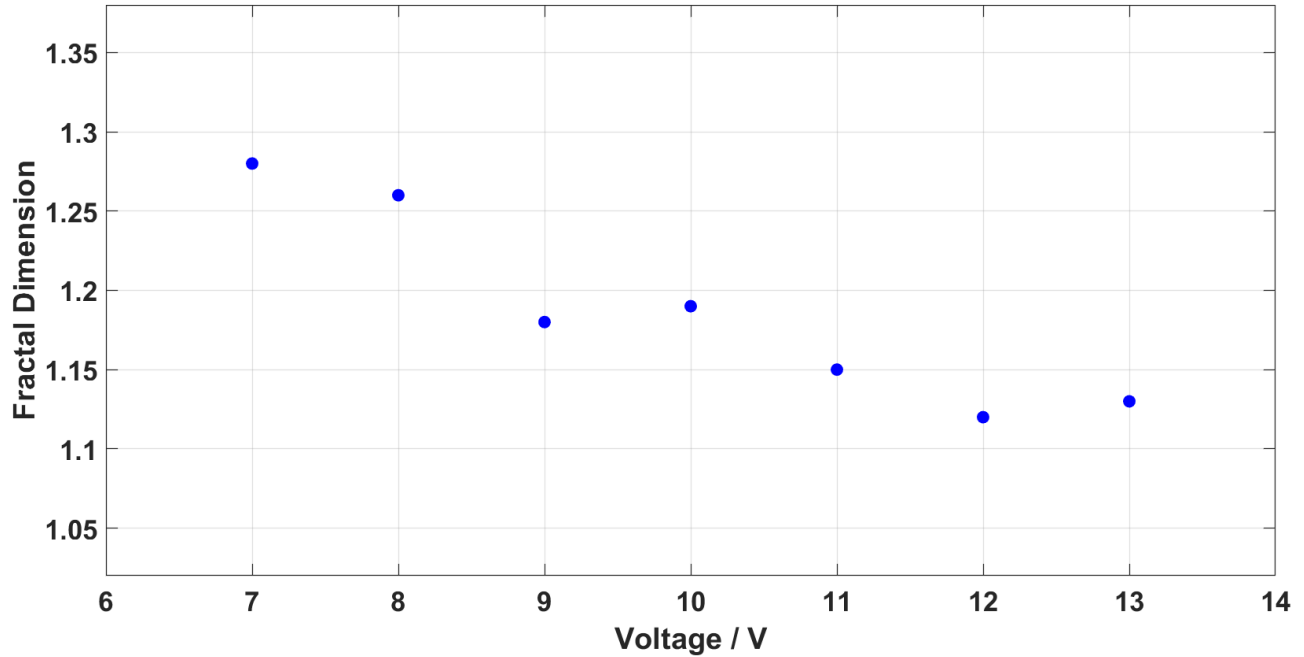

**Figure S2: Fractal dimension versus voltage**. Fractal dimension quantifies the complexity of a pattern, with a value of 1 corresponding to a straight line and 2 to a filled area; intermediate values indicate varying degrees of structural roughness. In this work, diffusion-limited aggregation causes filaments to exhibit fractal characteristics across multiple optical planes along the z-axis, though for this analysis the dimension was calculated from a top-view static image. The results show that increasing the applied voltage drives the filament's fractal dimension after connection closer to 1, indicating reduced structural complexity.

**Fractal dimension calculation using the box counting method**

To calculate the fractal dimension of a filament, binary filament images (white = filament, black = background) were generated from the original light-microscopy images and analyzed with the box-counting method. In this method each image is partitioned into a grid of non-overlapping boxes of side length $s$. The grid size is decreased by varying $s$, with the box count $N(s)$ being recorded for each grid.[1] The box count $N(s)$ was computed as the number of boxes occupied by the filament, where a box is considered occupied if at least one pixel in the box belongs to the filament. The fractal dimension $D$ was estimated as the negative slope of the linear regression of $\log(N(s))$ versus $\log(s)$, with K being a constant:[1]

$$\log(N(s)) = \log(K) - D \log(s)$$

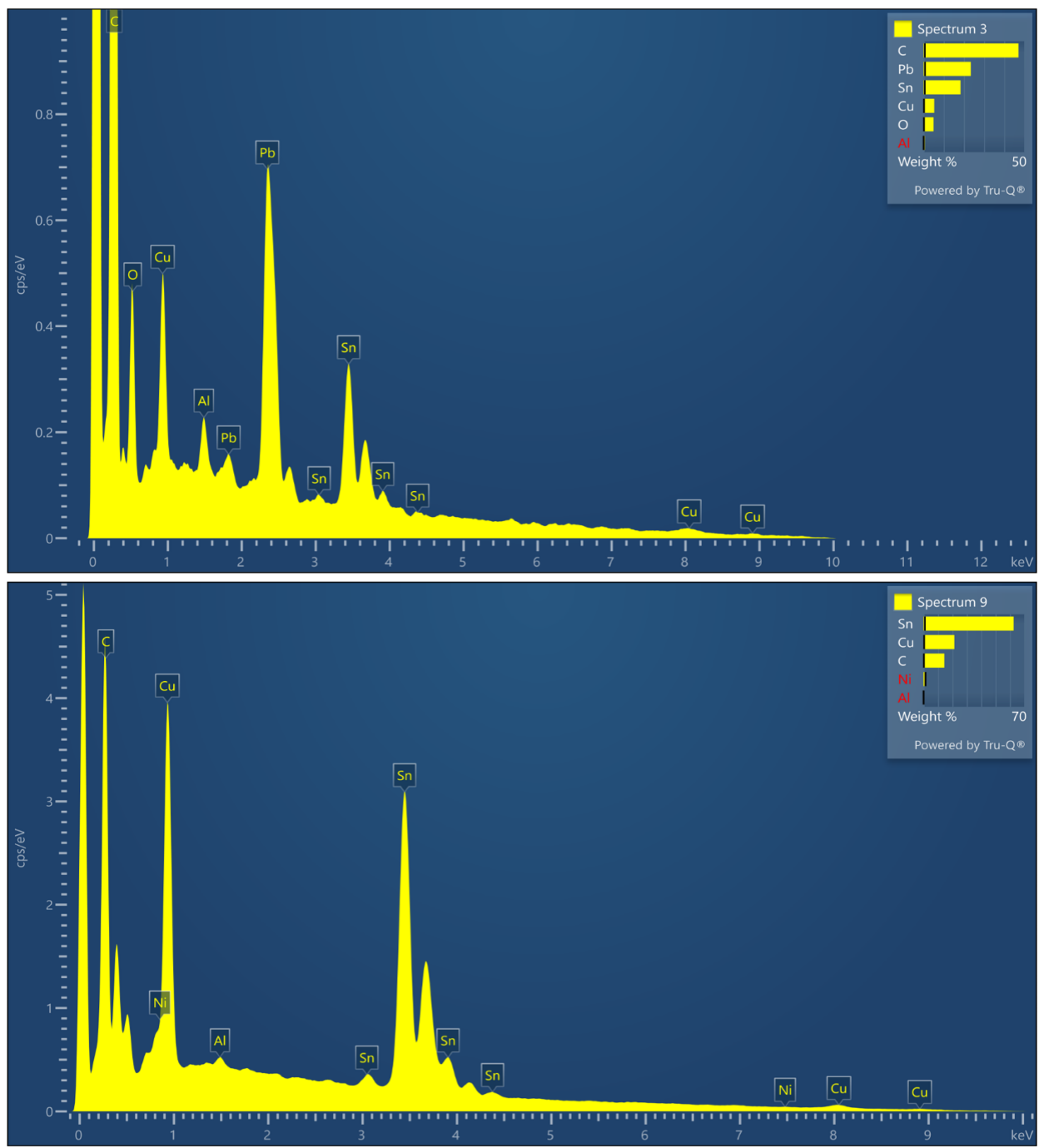

**Figure S3**: **EDX measurements of the surface of the virgin triangular electrodes**. For the experiments in the scope of this manuscript, surface coating without Pb was used. Due to the presence of Sn in the protective coating layer of the electrodes, the ionotronic system would contain both Cu and Sn in different percentages.

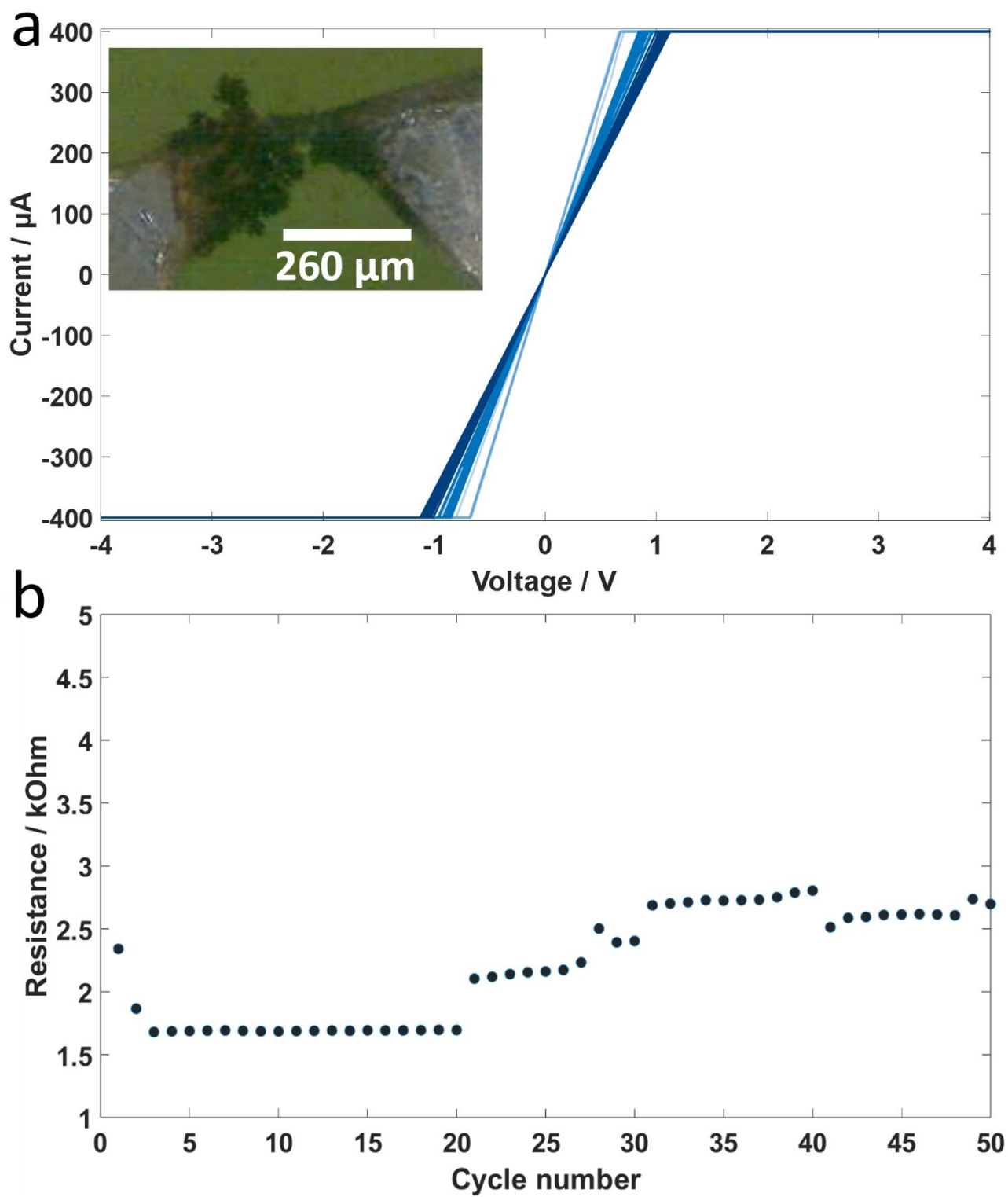

**Figure S4: Resistance of a filament of high thickness**. (a) Current versus voltage plot for 50 cycles during voltage sweeps of -4 to +4 over approximately 3 hours. 10 sets of cycles were recorded at each time window ranging from light blue to dark blue over time. A microscopic image of the filament is shown in the inset of the plot. (b) Resistance versus cycle number for the

50 voltage sweeps. The resistance remained stable at 1.6 kΩ for the first 20 cycles which were recorded in the first hour after the growth of the filament. With time the resistance of the filament steadily seems to increase. The proposed hypothesis is that over time, an oxide form in the filament which gradually increases the resistance. From cycle 20 to 21 there is a time gap of 66 minutes which contributes to the prompt increase in resistance. From cycle 30 to 31 there is a time gap of 25 minutes. Hence a comparatively smaller increase in resistance.

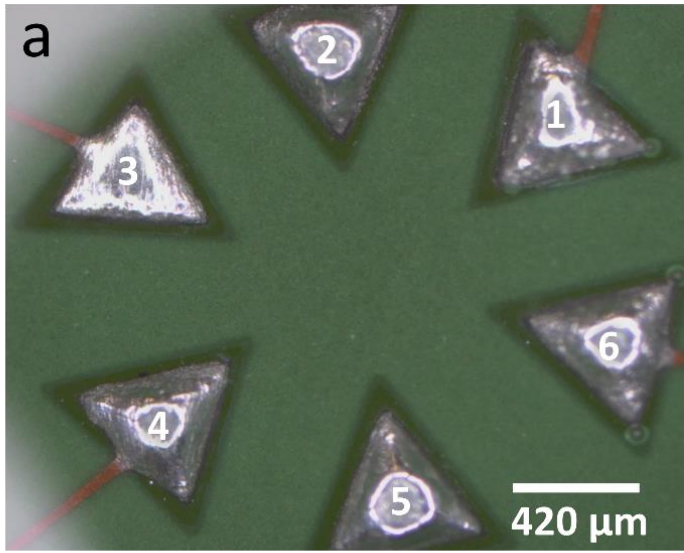 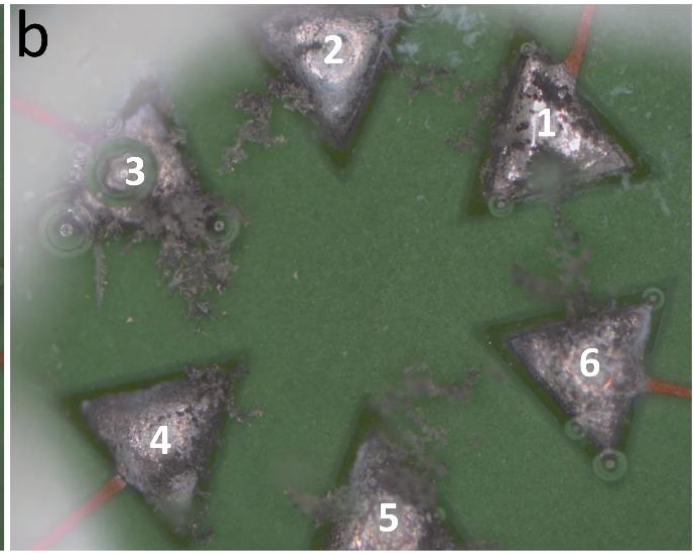
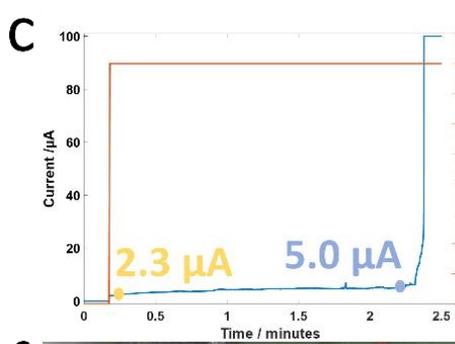 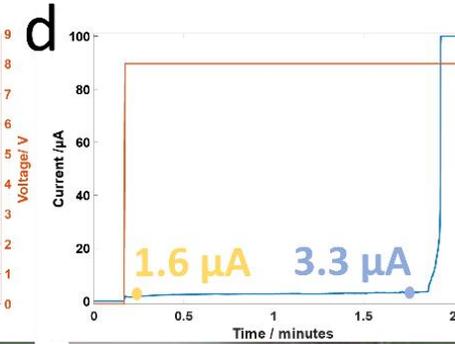 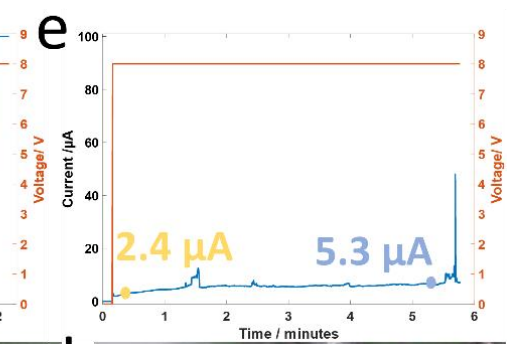
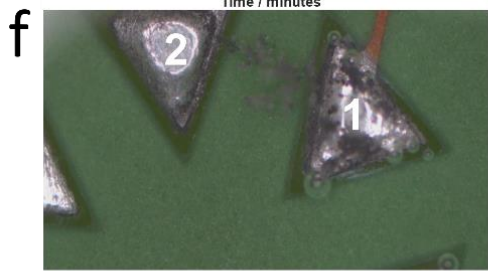 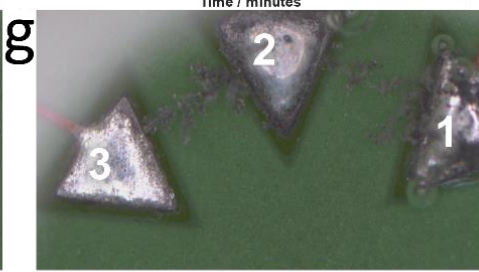 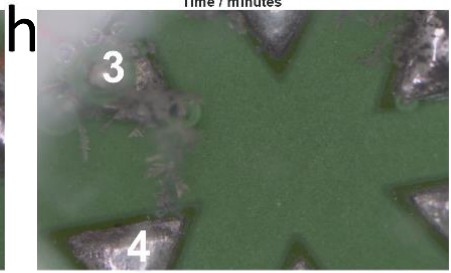
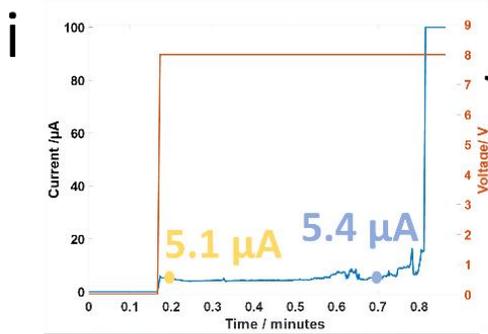 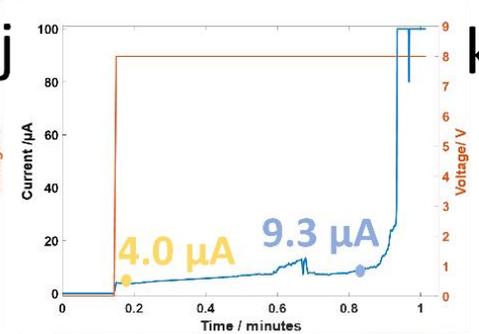 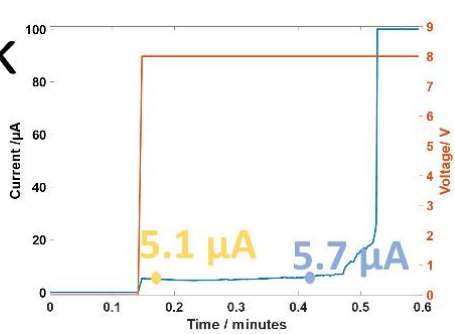
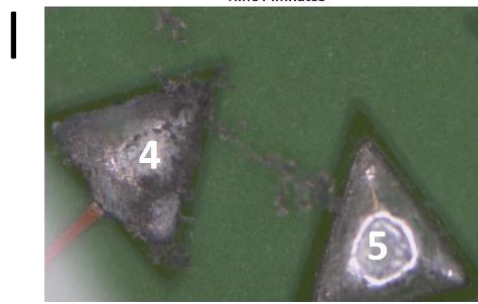 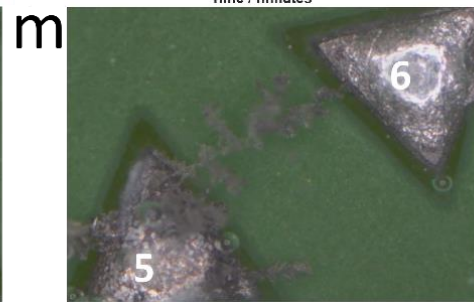 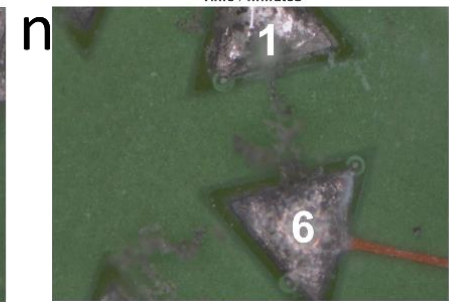

**Figure S5: Filament formation and breakage in a six-electrode system**. Microscope images a and b at the top of the figure shows the system before and after the experiment. Here two electrodes were individually connected in a step-by-step sequence. The current versus time profile for each experiment is shown in the bottom plots. Firstly, the electrodes 1 and 2 were connected under 8 V. A complete electrical connection was formed after 2 minutes with the current reaching compliance which was set to 0.1 mA. Secondly, the electrodes 2 and 3 were connected followed by electrodes 3 and 4, 4 and 5, 5 and 6, and finally 6 and 1 where a complete connection was formed in less than 30 seconds. At the end of the experiment, it was observed that some of the initial connections are broken due to them being inactive (image b).

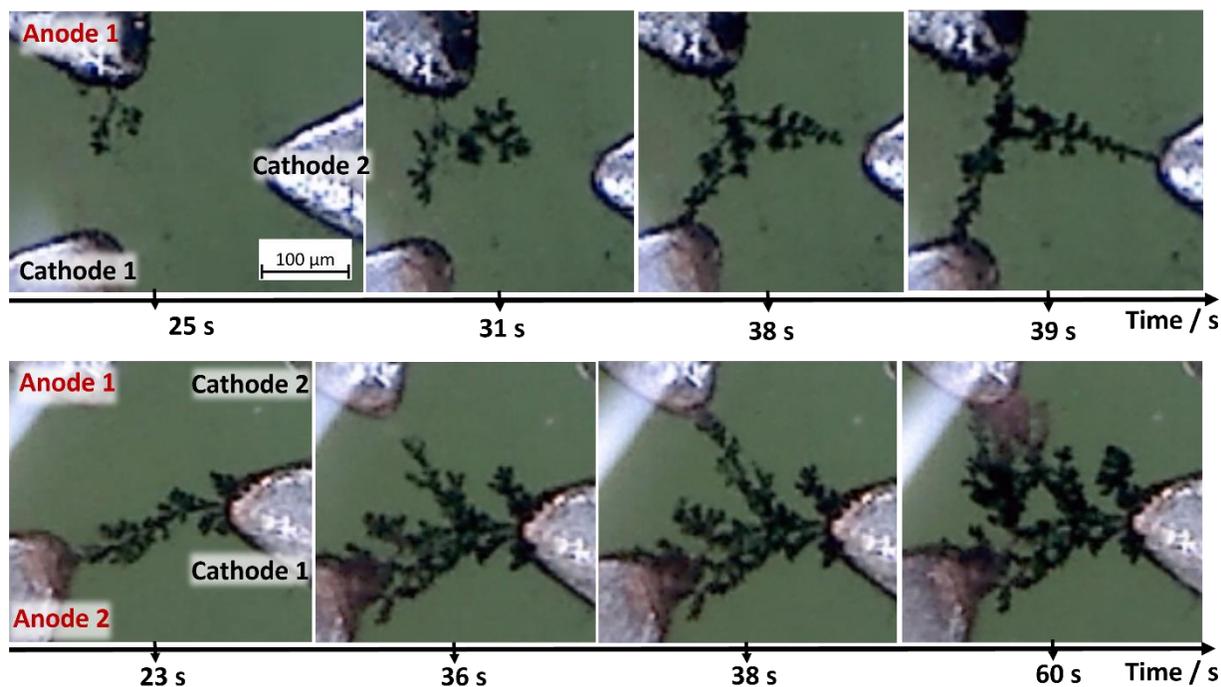

**Figure S6: Filament manipulation with electric field.** The two sets of images (top and bottom lines) show different ways of filament manipulation via an electric field. In the top images the filament is simultaneously connecting all 3 electrodes whereas in the bottom images firstly, a connection is made between cathode 1 and anode 2 and then the filament extends to anode 1. Top set of images: Anode 1 was grounded and respectively a voltage of -10 V and -8 V was applied to the cathode 1 and 2. Bottom set of images: the electrodes anode 1 and cathode 1 were connected together and anode 2 and cathode 2 (not visible in the images) were connected. At first 3 V is applied between anode 1 and cathode 1 where the cathode 1 was grounded and simultaneously 6 V was applied between anode 2 and cathode 2. After the connection was formed between anode 2 and cathode 1 the 6 V was reduced to 0 V so that a connection to anode 1 was made at only 3 V.

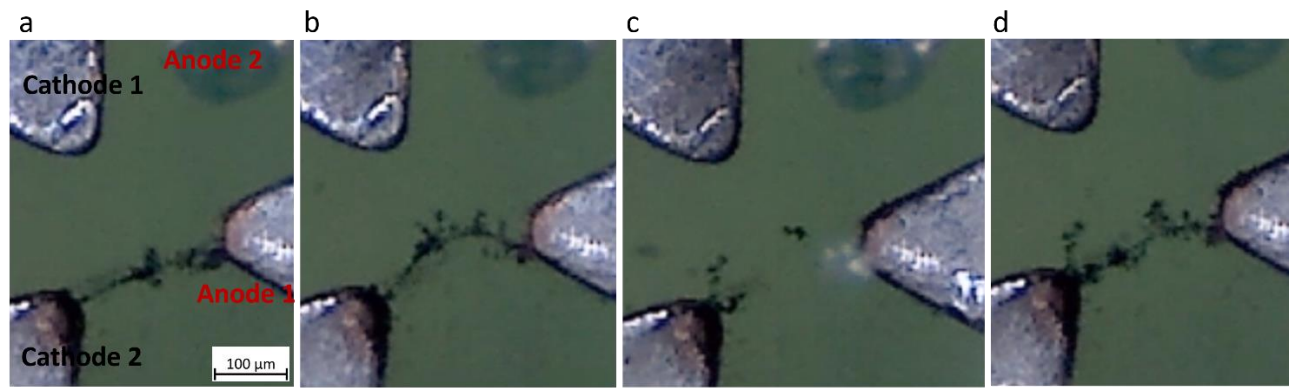

**Figure S7: Filament formation, breakage and reformation under the influence of an electric field**. 5 V was applied between anode 1 and cathode 1 and 10 V was applied between anode 2 and cathode 2. 5 s after the voltages were applied a filament was grown between anode 1 and cathode 2 as shown in the microscopic image a. Immediately after the 5 V was disconnected the filament bent in the direction of cathode 1 subsequently breaking the filament and reforming the filament within 15 s as shown in the microscopic images b, c, and d.

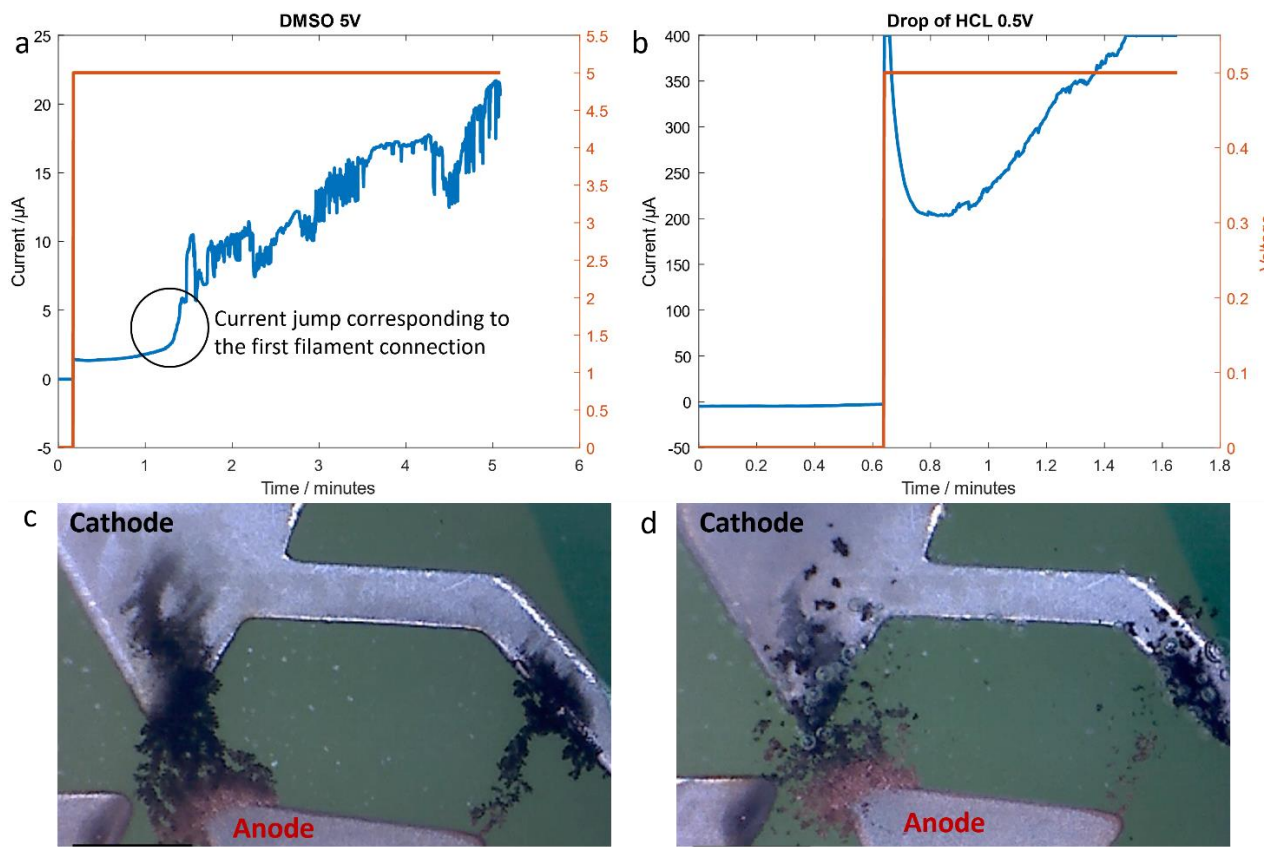

**Figure S8**: **Chemistry-based dissolving of the filament.** This figure shows images of a filament before and after 1 M HCl is added on it. The filament was grown in DMSO and the current versus time plot is shown on the left side of the figure together with an image of the final filament. This filament was grown with 5 V and the final resulting current of the filament was around 20 µA. Then 1 M HCl was added onto the already grown filament. The added HCl reacts with the oxide shell. This interrupts the connection depending on the thickness of the filament. the thicker filament in the image is still connected while the thinner filament is almost completely dissolved. The right image and the plot are after addition of HCl.

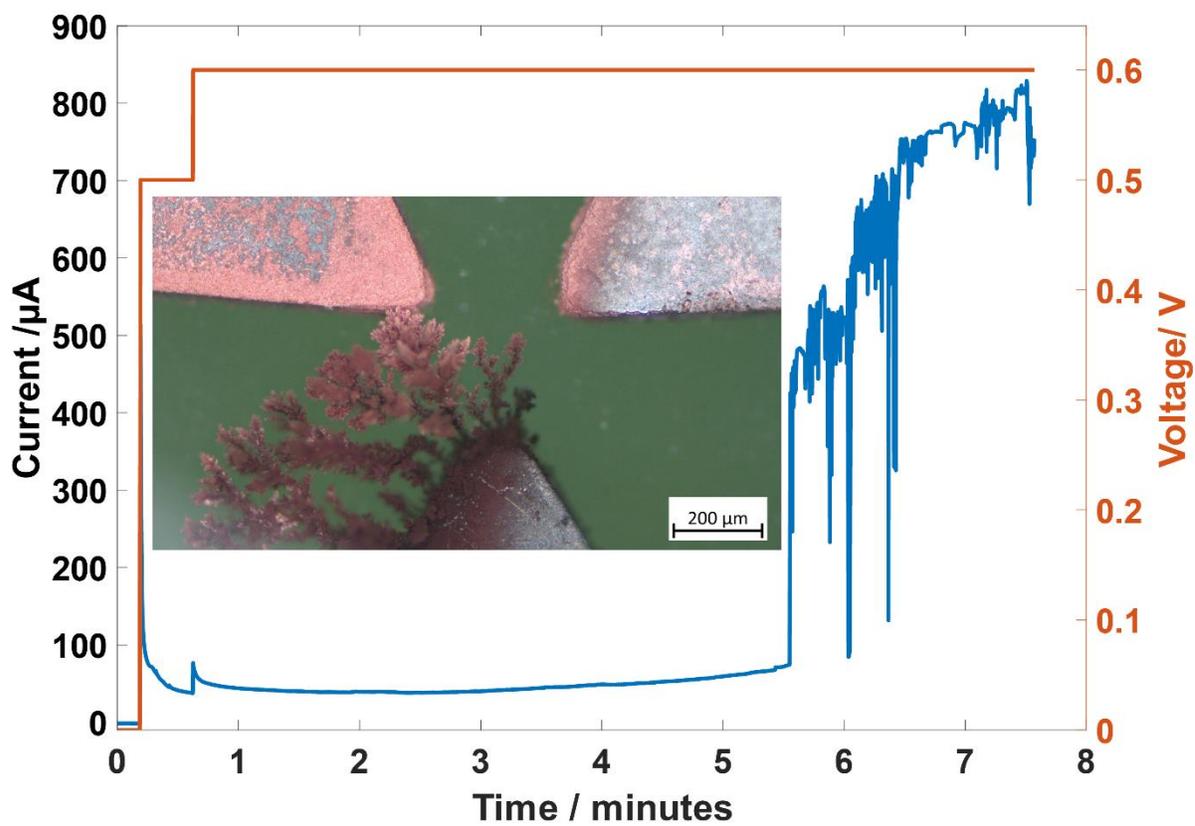

**Figure S9: Filament growth in 1 M HCl.** The current and voltage versus time plot here corresponds to the growth profile of the filament shown in the figure inset.

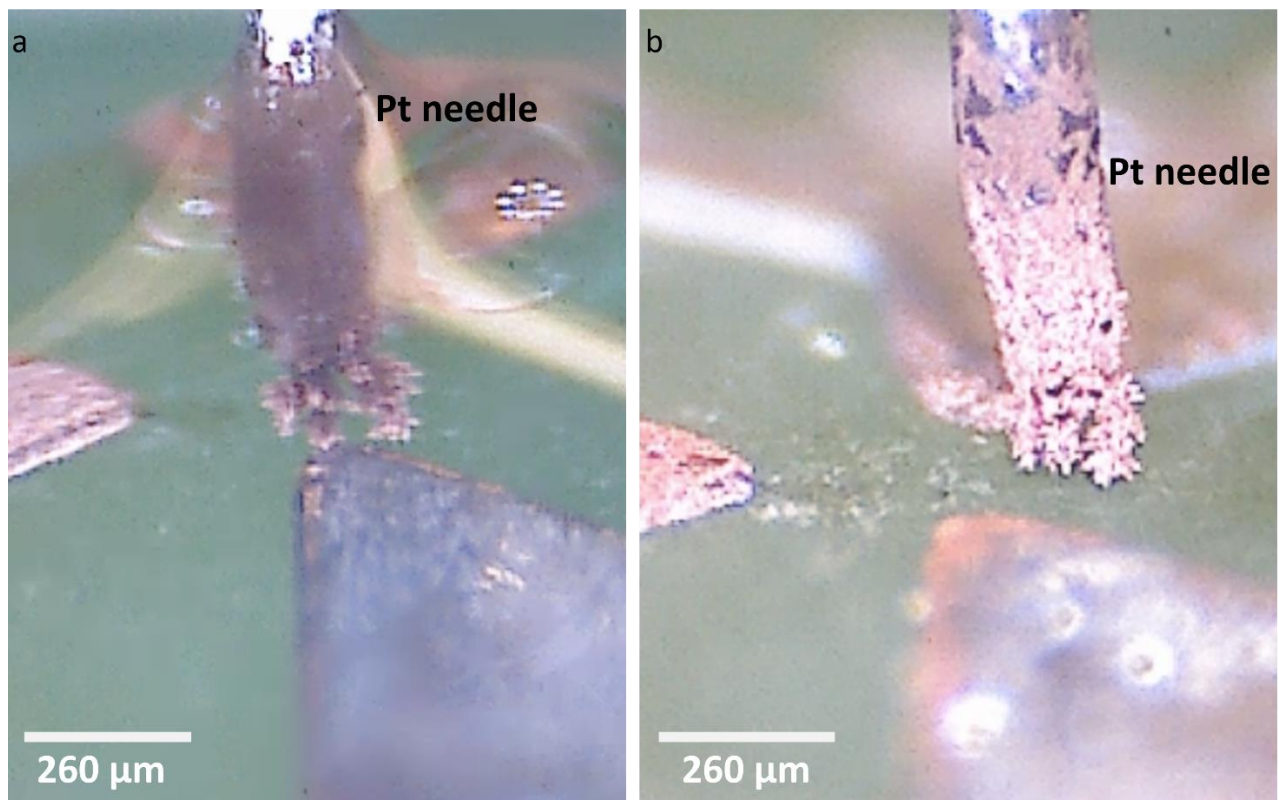

**Figure S10: Vertical filament growth in 1 M HCl**. The microscopic image on the left shows a filament grown between the triangular electrode and the Pt needle under 0.5 V in 1 M HCl. Here the copper electrode undergoes oxidation (anodic dissolution) and the filament grows from the Pt needle towards the triangular electrode. The image on the right shows the same filament after around 15 minutes under 0.05 V. The filament seems to have dissolved at the anode indicating that the applied voltage has not been sufficient enough to retain the filament.

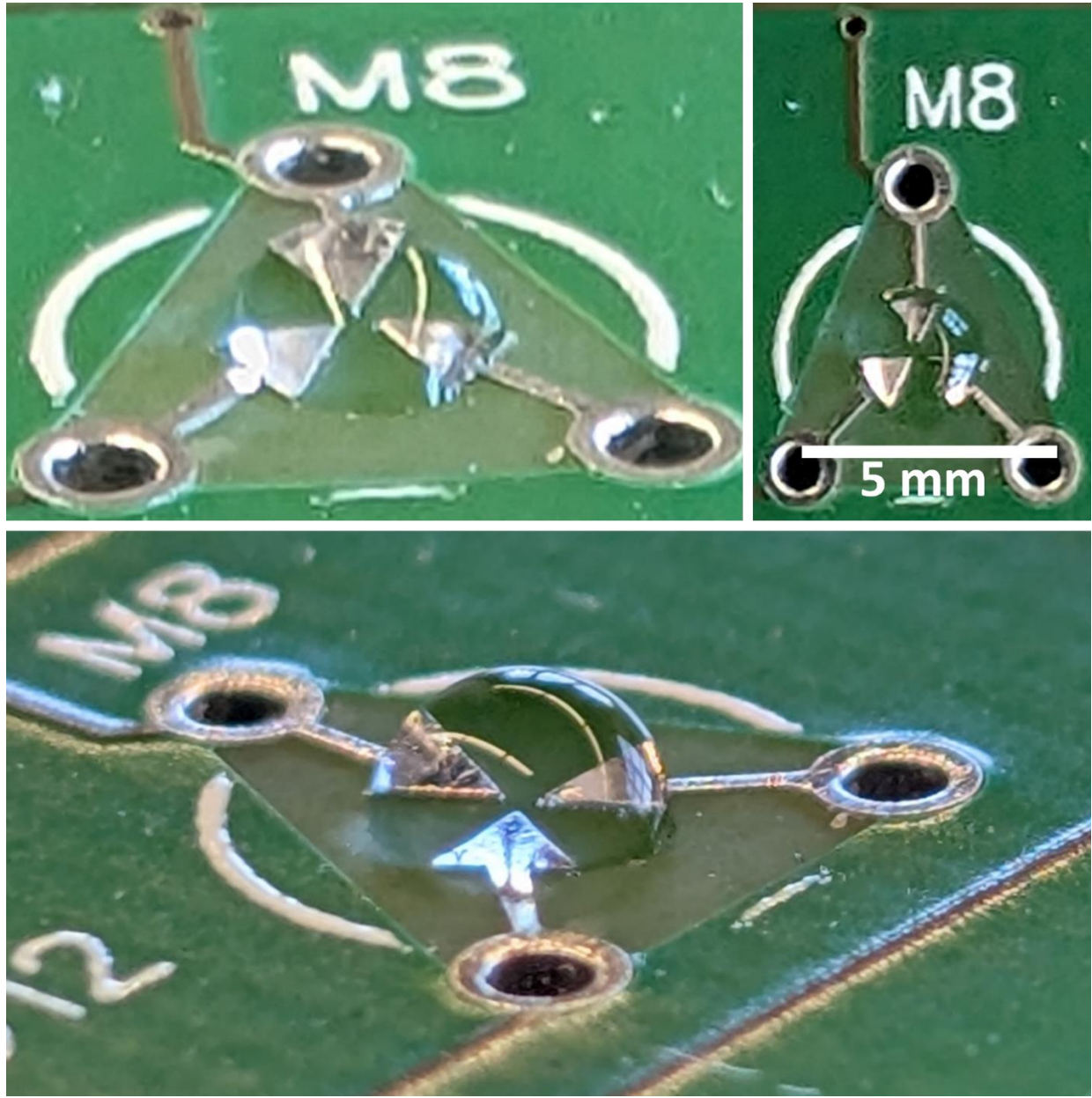

**Figure S11**: **Photos of the experimental setup**. The photos show The PCB in green and the three triangular electrodes with a 260 µm separation with a 5 µl liquid droplet.